\documentclass[epsfig,graphics,twocolumn,floatfix,mathbbm,prb]{revtex4}
\usepackage{graphicx}

\begin{document}

\title[a]{Coupling of an erbium spin ensemble to a superconducting resonator}

\author{Matthias U. Staudt$^1$}
\email{staudt@chalmers.se}
\author{Io-Chun Hoi$^1$}
\author{Philip Krantz$^1$}
\author{ Martin Sandberg$^1$}
\author{ Micha\"{e}l Simoen$^1$}
\author{ Pavel Bushev $^2$}
\author{Nicolas Sangouard$^3$}
\author{Mikael Afzelius$^3$}
\author{ Vitaly S. Shumeiko $^1$}
\author{G\"{o}ran Johansson$^1$}
\author{Per Delsing$^1$}
\author{C. M. Wilson$^1$}

\address{$^1$Microtechnology and Nanoscience, MC2, Chalmers University of Technology, SE-41296 G\"{o}teborg, Sweden}
\address{$^2$Physikalisches Institut, Karlsruhe Institute of Technology, D-76128 Karlsruhe, Germany}

\address{$^3$Group of Applied Physics, University of Geneva, CH-1211 Geneva 4, Switzerland}

\date{\today}

\begin{abstract}

A quantum coherent interface between optical and microwave photons can be used as a basic building block within a future quantum information network. The interface is envisioned as an ensemble of rare-earth ions coupled to a superconducting resonator, allowing for coherent transfer between optical and microwave photons. Towards this end, we have realized a hybrid device coupling a Er$^{3+}$ doped Y$_2$SiO$_5$ crystal in a superconducting coplanar waveguide cavity. We observe a collective spin coupling of 4 MHz and a spin linewdith of down to 75 MHz.
 
\end{abstract}

\maketitle

\section{Introduction}

The boundary set by the largest possible size of an individual quantum processing unit can be surpassed by a quantum network \cite{Kimble08}. Thus quantum networks may be crucial for the applications of quantum computing, communication and metrology \cite{Nielsen2000, Gisin2007, Giovanetti2004}. In such networks, quantum information is distributed through the network and processed in nodes \cite{Hoi2011}.  Entanglement is distributed via the channels, where optical photons carry the quantum information. The nodes can contain ensembles of real or artificial ions or atoms. Superconducting qubits \cite{Wallraff2004} are promising solid-state candidates for such nodes. As superconducting qubits work in the microwave range (GHz) and optical photons are typically in the telecom frequency (200 THz), a coherent quantum interface that can bridge the gap between these energy regimes is needed. This would lead to the combination of two fruitful research areas: quantum communication using optical photons and quantum processing using superconducting quantum circuits.  

Strong coupling of a superconducting qubit to a highquality superconducting transmission line resonator has been demonstrated \cite{Wallraff2004}, allowing for cavity quantum electrodynamic experiments on a chip (circuit QED). This achievement has enabled a number of quantum optics experiments using superconducting circuits \cite{Wilson2007, sandberg2008, Johansson2009, Wilson2010a, Wilson2010b, Wilson2010}. For instance, the dynamical Casimir effect\cite{Wilson2011}, number-resolving photon detection \cite{Schuster2007}, single-photon generation \cite{Houck2007}, two-qubit coupling via the circuit cavity \cite{Majer2007} and three - qubit entanglement \cite{Dicarlo2011, Neeley2011} have been demonstrated. Recently superconducting qubits having coherence times on the order of several tens of $\mu$s have been realized \cite{Paik2011}. 

Hybrid proposals \cite{sorenson2004,Imamoglu2009,Verdu2009,Wallquist2009} aim at integrating various types of quantum systems in a circuit QED setting, for instance, atomic ensembles that would interact with the microwave resonator via electric dipole transitions (e.g. Rydberg atoms or polar molecules\cite{andre2006}) or via magnetic spin transitions (alkali atoms, silicon spins). The latter are particularly challenging since the magnetic dipole transitions are very weak. Even in a strongly confined resonator volume, the single-spin coupling rate would be far from sufficient, typically $g_{c}/2\pi\sim$ 10-100 Hz \cite{Schoelkopf2008}. The coupling can be significantly increased \cite{sorenson2004,Imamoglu2009,Verdu2009,Wallquist2009} via a collective interaction with a spin ensemble, where the collective coupling scales as $g_{coll}=\sqrt{N}g_{c}$ with $N$ the number of atoms. This approach has recently attracted much attention and led to experimental realizations coupling superconducting resonators \cite{Kubo2010,Schuster2010,Kubo2011,am2011,Bushev2011} to nitrogen-vacancy centers and Er$^{3+}$ ions. 

In this paper, we investigate experimentally the coupling between an ensemble of Er$^{3+}$ ions doped into a Y$_2$SiO$_5$  host crystal and a superconducting coplanar waveguide resonator (CPWR) at mK temperatures. For a future optical to microwave interface, Er$^{3+}$ ions are attractive because of the $^4I_{15/2}$  $\rightarrow$ {$^4I_{13/2}$ transition around 1540 nm, which is the so-called telecom window where absorption losses are minimal in optical fibers. Its strong first-order Zeeman effect allows for easy tuning of the spin resonance to the $\sim$  5 GHz operation frequency of circuit QED experiments. This alleviates the need for strong magnetic fields, which are incompatible with superconductivity. Furthermore, an optical quantum memory has recently already been demonstrated \cite{Lauritzen10} at this wavelength.

The Er$^{3+}$ ion replaces Y$^{3+}$ ions in the in the Y$_2$SiO$_5$  host crystal, which exists in two crystallographically inequivalent sites having C1 symmetry (hereafter named sites 1 and 2). In this work, we investigate spectroscopically both sites down to temperatures of 50 mK. Recently, magnetic coupling of Er$^{3+}$ ions to a superconducting cavity has been shown in a parallel work for one occupied site and down to a temperature of 280 mK \cite{Bushev2011}. To the best of our knowledge, the spectroscopic measurements presented here are done at the lowest temperature rare-earth ions have been spectroscopically investigated in a crystalline host matrix, though temperatures of lower than 100 mK have investigated already in an amorphous host matrix\cite{Hegarty1983, Staudt2006}. 

\section{Theory}

Rare-earth (RE) ions doped into inorganic crystals are interesting quantum systems owing to their long optical and spin coherence times\cite{Mac2002,tittel10,sang2011}. They have recently been considered as qubits for quantum computing \cite{longdell2004,rippe2008} and as quantum memories \cite{sab2010,hedges10, sag11, clausen11, Damon10} for optical photons in the context of quantum repeaters for long-distance communication. 

RE ions can be divided into two groups; non-Kramers and Kramers ions, with an even and odd number of 4$f$ electrons, respectively. For the former type, the interaction with the typical low-symmetry crystal environment usually produce a closely spaced (10 - 100 MHz) hyperfine manifold in the ground state. As a result, the non-Kramers ions generally do not have resonances appropriate for circuit QED experiments. The Kramers ions, \emph{e.g.} Erbium, on the other hand, are, in low-symmetry environments left, with two degenerate electronic spin levels resulting in effective spin $S$=1/2 systems, named Kramers doublets. The magnetic moment of these Zeeman states can, for a given direction of the magnetic field, be expressed as $\mu_m=g\mu_b$ where the angle - dependent $g$ - factor is typically in the range 1-10 and $\mu_b$ is the Bohr magneton, corresponding to magnetic tuning factors of 14-140 GHz/Tesla. Therefore a sub-Tesla magnetic field would generally be sufficient to tune the spin transition into resonance with the CPWR. This approach is chosen here and has the advantage that RE spin ensembles have been studied in this frequency range for decades using electron paramagnetic resonance (EPR) experiments \cite{GuillotNoel2006}. 

The coupling rate of a single spin to the CPWR cavity mode is given by \cite{Imamoglu2009} $g_c=\mu_m\sqrt{\mu_0 \omega_r/(2\hbar V_c)}$, where $V_c$ denotes the cavity-mode volume, $\omega_r$ the cavity frequency and $\mu_0$ the vacuum permeability. This results in an ion coupling rate of  the order of Hz for a typical electron spin magnetic moment. The single spin-CPWR coupling is thus insufficient to overcome the dissipation rate $\kappa\sim 0.1-1$ MHz typical for a CPWR. The strong coupling regime can, however, be reached via an enhanced collective coupling of a spin ensemble to the CPWR \cite{Imamoglu2009, Kubo2010, am2011}: $g_{coll}=g_{c}\sqrt{N}$. The range in the number of spins $N$ for reaching strong collective coupling, \emph{i. e.} : $g_{coll}>\kappa$,  is of the order $N>\kappa^2/g_c^2\sim10^6-10^{10}$. Diniz \cite{Diniz2011} \emph{et al.} showed recently that increasing the collective coupling suppresses the effects of inhomogeneous broadening, as long as the emitters spectral distribution decreases quicker than $\omega^{-2}$. 

The hyperfine interaction in the $^4I_{15/2}$ ground-state has been studied by O. Guillot-No\"{e}l \emph{et al.} using EPR \cite{GuillotNoel2006}. For the even isotopes $^{162, 164, 166, 168, 170}$Er$^{3+}$ the interaction can be described by a spin Hamiltonian consisting of a single Zeeman interaction term: 

\begin{equation}H=\mu_b\mathbf{B \cdot g \cdot S,}
\label{Hspin}
\end{equation}
while in the case of the only odd isotope $^{167}$Er$^{3+}$ hyperfine and quadrupole interaction terms must be added: 
\begin{equation}H=\mu_b\mathbf{B \cdot g \cdot S}+\mathbf{I \cdot A \cdot S}+\mathbf{I \cdot Q \cdot I}
\label{Hspin}
\end{equation}

Here $\mathbf{B}$ denotes the external magnetic field, $\mathbf{S}$ the electronic spin and $\mathbf{I}$ the nuclear spin. The matrices $\mathbf{g}, \mathbf{A}$ and $\mathbf{Q}$ are the matrices describing the electronic Zeeman, hyperfine and quadrupole interactions. We neglect the weaker nuclear Zeeman effect. These are generally anisotropic due to the interaction with the surrounding crystal environment. Due to the low-symmetry sites in Y$_2$SiO$_5$, this leads to very strong spin mixing, resulting in weaker transition rules for the spin transitions in $^{167}$Er$^{3+}$.

The Er$^{3+}$ ions occupy two crystallographically inequivalent sites in the Y$_2$SiO$_5$  host crystal. Each of these sites has two magnetically inequivalent subclasses that are related by C2 symmetry. Thus, depending on the magnetic field orientation, up to four Zeeman transitions are expected for the even isotopes\cite{Sun08}.

\section{Experimental Setup}

We couple the Er$^{3+}$ ions to a $\lambda/2$ CPWR, having a Q value of 568.
The CPWR is realized by a 120 nm film of Niobium sputtered on a silicon substrate (380 $\mu m$) having a center strip with a width of 100 $\mu$m and gaps of 70 $\mu$m to the ground planes. 

Integration of the RE-doped crystal is done by placing the crystal directly on the superconducting cavity with pressure applied by a teflon frame. The cavity has its resonance frequency at 4.4 GHz. The crystal is doped with 0.02 $\%$ Er$^{3+}$ ions.

 We have measured the sample down to $\sim$ 50 mK in a dilution refrigerator equipped with a superconducting magnet. The external magnetic field is applied parallel to the surface of the superconducting chip and is ramped at 3.3 mT/s. The crystal has a dimension of 3.5$\times$4$\times$3 mm$^3$ and is cut along the optical extinction axes D$_1$, D$_2$ and b. The magnetic field is applied both within the b - D$_1$ plane and the superconducting chip surface and is parallel to the b-axis. The D$_2$-axis is perpendicular to this plane. We measure the microwave transmission coefficient, S$_{21}$, which contains both magnitude and phase, as a function of magnetic field. The microwave power at the entrance of the cavity is about -110 dBm.   

\section{Results}

We sweep the external magnetic field between 0 mT and 200 mT. For each applied magnetic field we measure S$_{21}$ (see Fig.~\ref{fig1}) and extract the Full Width at Half Maximum (FWHM) of the cavity resonance peaks. The extracted FWHM is plotted as a function of magnetic field in Fig.~\ref{lowfield} and Fig.~\ref{highfield}. The S$_{21}$ spectrum shown in Fig.~\ref{fig1} is for the case where the Er$^{3+}$ - ions are off - resonance. When the Er$^{3+}$ - ions are tuned into resonance with the cavity mode, we expect to observe an increase in the measured cavity linewidth. This is beacause the resonant ions provide a loss channel for the microwave photons. Two different regions of applied field are covered in higher resolution, one around 40 mT (Fig.~\ref{lowfield}) and the other around 140 mT (Fig.~\ref{highfield}). In each magnetic field region, a pair of peaks are visible. As mentioned, depending on the magnetic field orientation, up to four transitions are expected for Erbium, which is the case observed here. The pair of lines originating from the two crystallographic sites we denote as group 1 and 2, respectively. Lines within each pair, which are due to the magnetically inequivalent subclasses, are denoted as lines a and b. We extract the g-factor from the peak position for both sites (see table 1). 
For the same crystal orientation Guillot-No\"{e}l, et al. \cite{GuillotNoel2006} find a value for the g-factor of g$_{1}=8.92$ and g$_{2}=2.78$. The splitting for each transition and the deviation in g value we attribute to a slight misalignment of the external B field to the b-axis. 

If the magnetic field were perfectly aligned along the b-axis, then the subclasses would be magnetically equivalent and only two transitions visible. However, the splitting for both low and high field region are small and from the measurements in Refs. \cite{bottger02, Sun08} we can infer that the misalignment is of the order of $\pm 5^\circ$ compared to the b-axis.

\begin{figure} [tbp]

    \centering
    \
    \includegraphics[width=0.45\textwidth]{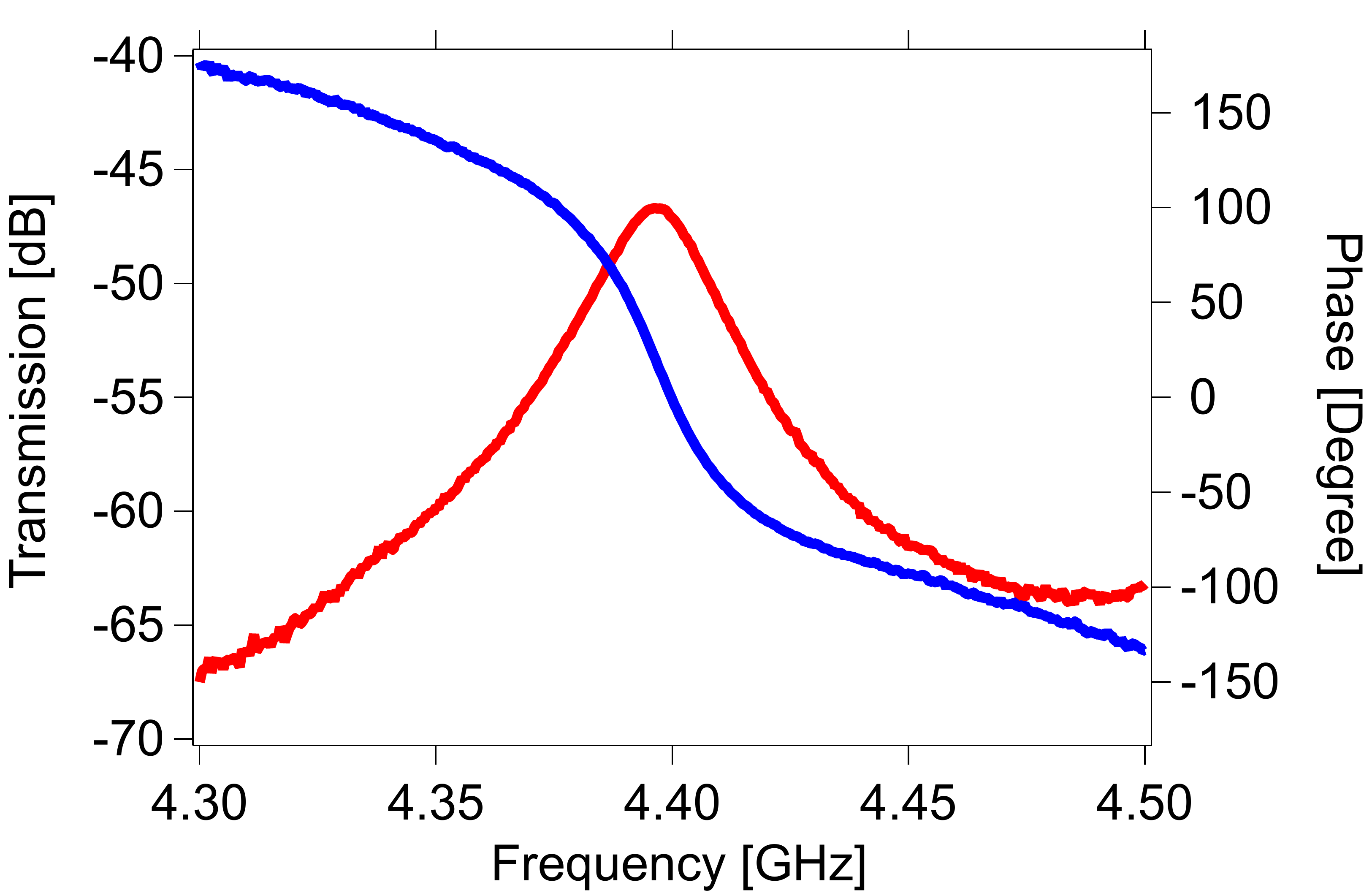}

    \caption{Transmission spectrum of a 4.4 GHz resonator. Both magnitude (red) and phase (blue) signal are shown. Losses in the lines to the resonator are about 90 dB including room-temperature and cold attenuators. The total gain is about 40 dB, giving rise to a maximum peak transmission of about -47 dB.}
    \label{fig1}
\end{figure}

\begin{figure} [tbp]
 \centering

  \includegraphics[width=0.5\textwidth]{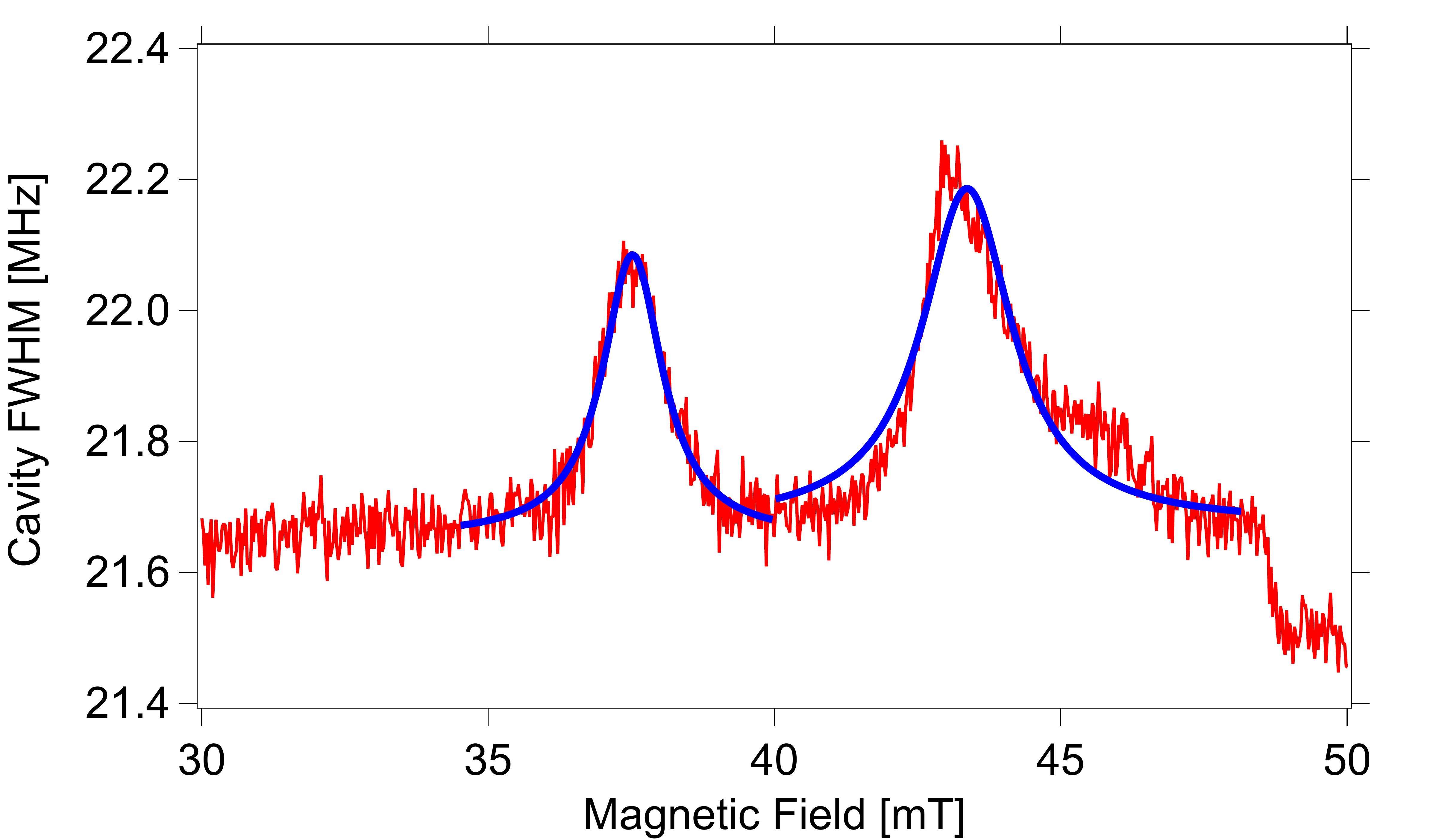}
  \caption{Resonator linewidth as a function of magnetic field in the field region around 40 mT.
Magnetic g-factors between 7 and 8 are extracted. Within each pair, the two lines represent spins aligned along different crystal axes that are related by symmetry. From the fit (solid lines, see text for details) we obtain a collective coupling rate of about 4 MHz and a spin linewidth of about 75 MHz.}
  \label{lowfield}
\end{figure}

We model both the spin ensemble and the cavity as a single-mode harmonic oscillator similar to Schuster \emph{et al.}$\cite{schuster10}$.
The detuning between the cavity resonance $\omega_r$ and atomic resonance due to the Zeeman splitting is: $\Delta_{Z}=\omega_r-g\mu_bB/\hbar$. 
In principal, the resonance is split due to hyperfine interaction for the $^{167}$Er$^{3+}$ ions \cite{GuillotNoel2006} (see also \cite{hyperfine}). As the hyperfine peaks are not visible in the observed spectrum, the hyperfine interaction is not taken into account in our fit. The total width is then given as $\Gamma_{\text{tot}}=\kappa+ \Gamma_{Z}$ where $\kappa$ is the cavity linewidth and $\Gamma_{Z}$ the spin induced linewidth.
The spin induced linewidth is:

\begin{equation}
\Gamma_{Z}=\frac{2g_{coll}^2\gamma}{(\Delta_{Z}^2+\gamma^2)} 
\end{equation}

where g$_c$ is the coupling constant.

For the pair of transitions having a maximum at (a) 37.7 mT and (b) 43 mT (site 1) and at (a) 125 mT and (b) 154 mT (site 2) we extract the parameters shown in table 1.

\begin{table}
\begin{tabular}{|c||c|c|c|c|}
\hline
Site & 1a & 1b & 2a & 2b \\
\hline
g & 8.37 & 7.25 & 2.51 & 2.04  \\
\hline
$\gamma$ [MHz] & $74.9$ & $96.6$ & $101$ & $136$ \\
\hline
g$_{coll}$ [MHz] & $4.02$ & $4.98$ & $6.07$ & $6.16$  \\
\hline
\end{tabular}
\caption{\label{Params} Measured parameters for sites 1 and 2.  Here g is the g-factor, $\gamma$ the spin linewidth and g$_{coll}$ the collective coupling constant.}
\end{table}

In order to estimate the collective coupling constant for a temperature approaching zero, we also perform a temperature dependent measurement (see Fig. \ref{temp} for the case of site 2b). The coupling constant g$_{coll}$ is measured between 70 mK and 500 mK. The population for the lower Zeeman level scales as $N_{1}\sim N \frac{ \exp{(-x)}}{\exp{(-x)}+\exp{(x)}}$ with $x=\hbar \omega/k_{b}T$  and N the total number of atoms. For the upper level we have $N_{2}\sim N \frac{ \exp{(x)}}{\exp{(-x)}+\exp{(x)}}$. The collective coupling $g_{coll}(T)$, depends on the temperature through the relative difference of these populations, namely: $g_{coll}(T)=g_{c}\sqrt {N\tanh(x)}=g_{coll}(0)\sqrt {\tanh( x)}$. Fixing the resonance frequency, we obtain a collective coupling constant of 6.14 MHz for site 2a extrapolating to zero temperature. From the fit we infer that the sample is highly polarized, reaching $90\%$ polarization for the lowest temperatures in our experiment.

\begin{figure} [t]

    \centering
    \
    \includegraphics[width=0.5\textwidth]{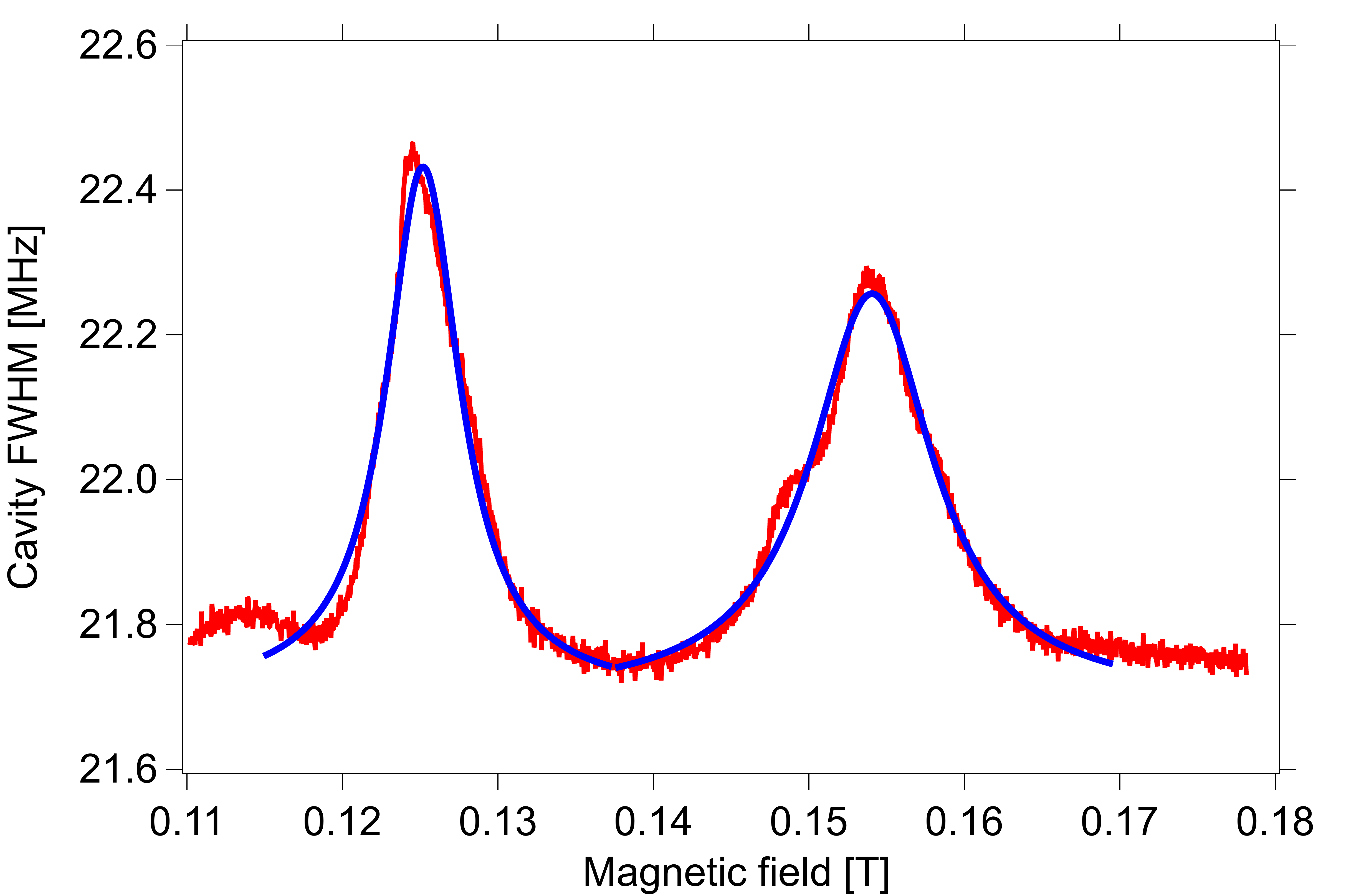}

    \caption{Resonator linewidth as a function of magnetic field in the field region around 140 mT.
Magnetic g-factors between 2 and 3 are extracted. Within each pair, the two lines represent spins aligned along different crystal axes that are related by symmetry.
From the fit (solid lines, see text for details) we obtain a collective coupling rate of about 6 MHz and a spin linewidth of about 100 MHz.}
    \label{highfield}
\end{figure}

\section{Conclusions}

In conclusion, we have shown coupling between an Er$^{3+}$ ensemble doped into a Y$_2$SiO$_5$ crystal and a superconducting CPW resonator, having a collective spin coupling of 4 MHz and a spin linewidth of down to 75 MHz, in good agreement with previous work\cite{Bushev2011}. The good coupling in the microwave regime, the widely used telecom transition and the good coherence properties show the potential for a coherent quantum optical - microwave interface made from an Er$^{3+}$ ensemble doped into a Y$_2$SiO$_5$ crystal.

Though the collective enhancement effect of the spins participating leads to a coupling in the MHz range, we are not yet entering the strong coupling regime. The usage of a high-Q cavity should be advantageous to reach that goal. Moreover a narrower inhomogeneous linewidth could be obtained by reducing the Er$^{3+}$ doping concentration. However, this will also imply less ion - spins participating, decreasing the coupling strength at the same time. Instead one could follow an interesting proposal\cite{Diniz2011} and increase the coupling strength by further increasing the Er$^{3+}$ doping concentration and enter a regime where the relaxation is governed by the single emitters properties taking advantage of so called "cavity protection."

\begin{figure} [t]

    \centering
    \
    \includegraphics[width=0.5\textwidth]{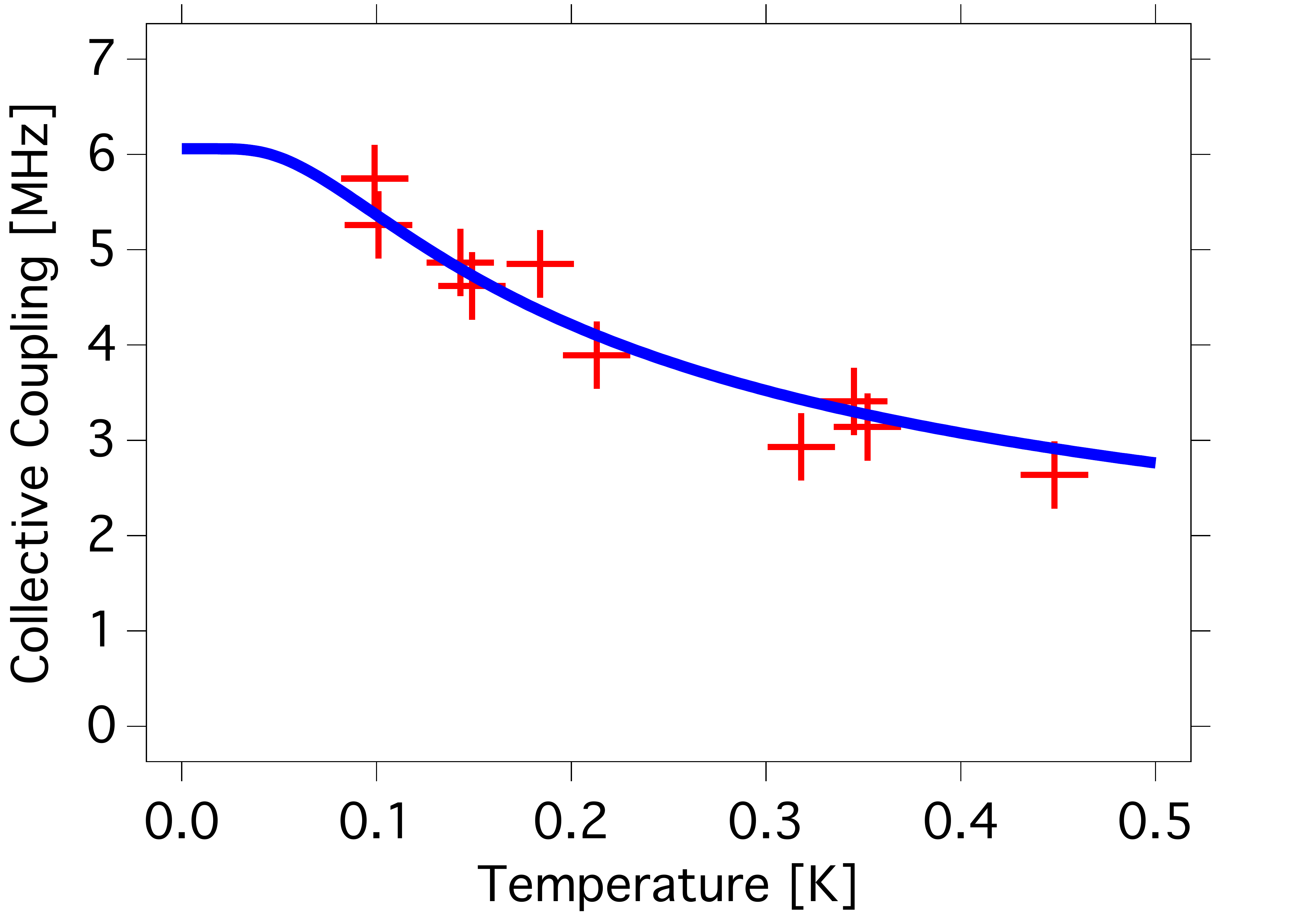}

    \caption{Temperature dependence of the coupling for site 2 having a magnetic g-factor of 2.04 (transition labelled 2b). We can explain the dependence by assuming that the number of spins participating in the ensemble decreases as the spins are thermally depolarized (see text for details). We see that we reach 90$\%$ polarization at the lowest temperatures.}
    \label{temp}
\end{figure}

For a future quantum-coherent interface, the low-field transition having a g - factor of 8.4 seems the most promising. Here a linewidth down to 75 MHz is obtained and the moderate external magnetic field does not suppress superconductivity in the CPWR.

Instead of using the Zeeman splitting, one could use the hyperfine structure present in Kramers ions with non-zero nuclear magnetic spin (i.e. for Erbium I=7/2). The hyperfine interaction for Erbium is in the range of several hundred MHz. This approach avoids the problem of inhomogeneity in the magnetic field, and transitions with low sensitivity to stray magnetic fields could be exploited. Although the properties of hyperfine transitions at zero or close to zero magnetic field has been less studied, we believe that this configuration, combined with the low-temperature conditions of circuit QED experiments, has advantages for the coherence properties of the spin ensemble.

\section{Acknowledgement}
We would like to thank F. Persson and M. Pierre for experimental help. This work was supported by the EU integrated project SOLID. P.B. acknowledges the financial support through BMBF project QUIMP and RiSC grant of KIT and MWK of Baden-Wuerttemberg. M.S., N.S. and M.A. acknowledge financial support through EU integrated project Q-Essence.

\bibliographystyle{unsrt}

\begin{thebibliography}{1}

\bibitem{Kimble08}
H. J. Kimble,
\newblock Nature \textbf{453}, 1028 (2008).

\bibitem{Nielsen2000}
M. A. Nielsen and I. L. Chuang,
\newblock Quantum Computation and Quantum Information, 
Cambridge University Press, Cambridge, England (2000).

\bibitem{Gisin2007}
N. Gisin and R. Thew,
\newblock Nature Photonics \textbf{1}, 165 (2007).

\bibitem{Giovanetti2004}
V. Giovannetti, S. Lloyd and L. Maccone, 
\newblock Science \textbf{306}, 1330 (2004).

\bibitem{Hoi2011}
I. Hoi, C. M. Wilson, G. Johansson, T. Palomaki, B. Peropadre, and P. Delsing,
\newblock Phys. Rev. Lett. \textbf{107}, 073601 (2011).

\bibitem{Wallraff2004}
A. Wallraff, D. I. Schuster, A. Blais, L. Frunzio, R.-S. Huang, J. Majer, S. Kumar, S. M. Girvin, and R. J. Schoelkopf,
\newblock Nature \textbf{431}, 162 (2004).

\bibitem{Wilson2007}
C. M. Wilson, T. Duty, F. Persson, M. Sandberg, G. Johansson, P. Delsing,
\newblock Phys. Rev. Lett. \textbf{98}, 257003 (2007).

\bibitem{sandberg2008}
M. Sandberg, C. M. Wilson, F. Persson, T. Bauch, G. Johansson, V. Shumeiko, T. Duty, and P. Delsing,
\newblock  Appl. Phys. Lett. \textbf{92} , 203501 (2008).

\bibitem{Johansson2009}
 J. R. Johansson, G. Johansson, C. M. Wilson, F. Nori, 
\newblock Phys. Rev. Lett. \textbf{103}, 147003 (2009).

\bibitem{Wilson2010a}
C.M. Wilson, G. Johansson, T. Duty, F. Persson, M. Sandberg and P. Delsing, 
\newblock Phys. Rev. B \textbf{81}, 024520 (2010).

\bibitem{Wilson2010b}
F. Persson, C. M. Wilson, M. Sandberg, G. Johansson and P. Delsing,
\newblock Nano Lett. \textbf{3}, 953 (2010).

\bibitem{Wilson2010}
C.M. Wilson, T. Duty, M. Sandberg, F. Persson, V. Shumeiko, P. Delsing, P,
\newblock Phys. Rev. Lett. \textbf{105}, 233907 (2010).

\bibitem{Wilson2011}
C. M. Wilson, G. Johansson, A. Pourkabirian, M. Simoen, J. R. Johansson, T. Duty, F. Nori and P. Delsing,
\newblock Nature \textbf{479}, 376 (2011).

\bibitem{Schuster2007}
D. I. Schuster, A. A. Houck, J. A. Schreier, A. Wallraff, J. M. Gambetta, A. Blais, L. Frunzio, J. Majer, B. Johnson, M. H. Devoret, S. M. Girvin, and R. J. Schoelkopf,
\newblock Nature \textbf{445}, 515 (2007).

\bibitem{Houck2007}
 A. A. Houck, D. I. Schuster, J. M. Gambetta, J. A. Schreier, B. R. Johnson, J. M. Chow, L. Frunzio, J. Majer, M. H. Devoret, S. M. Girvin, and R. J. Schoelkopf,
\newblock Nature \textbf{449}, 328 (2007).

\bibitem{Majer2007}
 J. Majer, J. M. Chow, J. M. Gambetta, J. Koch, B. R. Johnson, J. A. Schreier, L. Frunzio, D. I. Schuster, A. A. Houck, A. Wallraff, A. Blais,  M. H. Devoret, S. M. Girvin, and R. J. Schoelkopf,
\newblock Nature \textbf{449}, 443 (2007).

\bibitem{Dicarlo2011}
L. DiCarlo, M. D. Reed, L. Sun, B. R. Johnson,  J. M. Chow, J. M. Gambetta, L. Frunzio, S. M. Girvin, M. H. Devoret and R. J. Schoelkopf,
\newblock Nature \textbf{467}, 574 (2011).

\bibitem{Neeley2011}
M. Neeley, R. C. Bialczak, M. Lenander, E. Lucero, M. Mariantoni, A. D. O’Connell, D. Sank, H. Wang, M. Weides, J. Wenner, Y. Yin, T. Yamamoto, A. N. Cleland and J. M. Martinis,
\newblock Nature \textbf{467}, 570 (2011).


\bibitem{Paik2011}
H. Paik, D. I. Schuster, L. S. Bishop, G. Kirchmair, G. Catelani, A. P. Sears, B. R. Johnson, M. J. Reagor, L. Frunzio, L. I. Glazman, S. M. Girvin, M. H. Devoret, and R. J. Schoelkopf,
\newblock Phys. Rev. Lett. \textbf{107}, 240501 (2011).



\bibitem{sorenson2004}
A. .S. S\o{}rensen, C. H. van der Wal, L. I. Childress, M. D. Lukin, 
\newblock Phys. Rev. Lett. 92, 063601 (2004).

\bibitem{Imamoglu2009}
A. Imamo\v{g}lu,
\newblock Phys. Rev. Lett. \textbf{102}, 083602 (2009).

\bibitem{Verdu2009}
J. Verd\'{u}, H. Zoubi, Ch. Koller, J. Majer, H. Ritsch, and J. Schmiedmayer,
\newblock Phys. Rev. Lett \textbf{103}, 043603 (2009).

\bibitem{Wallquist2009}
M. Wallquist, K. Hammerer, P. Rabl, M. Lukin, and P Zoller,
\newblock Phys. Scr. T \textbf{137}, 014001 (2009).  

\bibitem{andre2006}
A. Andr\'e, D. DeMille, J. M. Doyle, M. D. Lukin, S. E. Maxwell, P. Rabl, R. J. Schoelkopf and P. Zoller,
\newblock Nature Physics \textbf{2}, 636 (2006).

\bibitem{Schoelkopf2008}
R. J. Schoelkopf and S. M. Girvin,
\newblock Nature \textbf{451}, 664 (2008).

\bibitem{Kubo2010}
Y. Kubo, F. R. Ong, P. Bertet, D. Vion,  V. Jacques, D. Zheng, A. Dr\'eau, J.-F. Roch, A. Auffeves, F. Jelezko, J. Wrachtrup, M.F. Barthe, P. Bergonzo, D. Esteve,
\newblock Phys. Rev. Lett. \textbf{105}, 140502 (2010).

\bibitem{Schuster2010}
D. I. Schuster, A. P. Sears, E. Ginossar, L. DiCarlo, L. Frunzio, J.J.L.Morton, H. Wu, G.A.D. Briggs, B.B. Buckley, D.D. Awschalom, R. J. Schoelkopf, 
\newblock Phys. Rev. Lett. \textbf{105},  140501 (2010).

\bibitem{Kubo2011}
Y. Kubo, C. Grezes, A. Dewes, T. Umeda, J. Isoya, H. Sumiya, N. Morishita, H. Abe, S. Onoda, T. Ohshima, V. Jacques, A. Dr\'eau,  J.-F. Roch, I. Diniz, A. Auffeves, D. Vion, D. Esteve, P. Bertet, 
\newblock Phys. Rev. Lett. \textbf{107},  220501 (2011).

\bibitem{am2011}
R. Amsuss, C. Koller, T. Nobauer, S. Putz, S. Rotter, K. Sandner, S. Schneider, M. Schrambock, G. Steinhauser, H. Ritsch, J. Schmiedmayer, J. Majer, 
\newblock Phys. Rev. Lett. \textbf{107}, 060502 (2011).

\bibitem{Bushev2011}
P. Bushev, A. K. Feofanov, H. Rotzinger, I. Protopopov, J. H. Cole, C. M. Wilson, G. Fischer, A. Lukashenko, and A. V. Ustinov,
\newblock Phys. Rev. B. \textbf{84}, 060501 (2011).


\bibitem{Lauritzen10}
B. Lauritzen, J. Minar, H. de Riedmatten, M. Afzelius, N. Sangouard, C. Simon, and N.Gisin,
\newblock Phys. Rev. Lett. \textbf{104}, 080502 (2010).





\bibitem{Hegarty1983}
J. Hegarty, M. M. Broer, B. Golding, J. R. Simpson, and J. B. MacChesney, 
\newblock Phys. Rev. Lett. \textbf{51}, 2033 (1983).

\bibitem{Staudt2006}
M. U. Staudt, S. R. Hastings-Simon, M. Afzelius, D. Jaccard, W. Tittel, N. Gisin, 
\newblock Optics Communications \textbf{266}, 720 (2006).

\bibitem{Mac2002}
R. M. Macfarlane,
\newblock Journal of Luminescence \textbf{100}, 1 (2002).

\bibitem{tittel10}
W. Tittel, M. Afzelius, T. Chaneli\`ere, R. L. Cone, S. Kr\"{o}ll, S. A. Moiseev and M. Sellars,
\newblock Laser and Photonics Reviews \textbf{4}, 244 (2010). 

\bibitem{sang2011}
N. Sangouard, C. Simon, H. de Riedmatten and N. Gisin,
\newblock Rev. Mod. Phys. \textbf{83}, 3 (2011).

\bibitem{longdell2004}
J.J. Longdell and M.J. Sellars,
\newblock Phys. Rev. A \textbf{69}, 032307 (2004).

\bibitem{rippe2008}
L. Rippe, B. Julsgaard, A. Walther, Y. Ying, S. Kr\"oll,
\newblock Phys. Rev. A \textbf{77}, 022307 (2008).

\bibitem{sab2010}
M. Sabooni,F. Beaudoin, A. Walther, N. Lin, A. Amari, M. Huang, S. Kr\"oll, 
\newblock Phys. Rev. Lett. \textbf{105}, 060501 (2010).


\bibitem{hedges10}
M. P. Hedges, J.J. Longdell, Y. Li and M.J. Sellars, 
\newblock Nature \textbf{465}, 1052 (2010).

\bibitem{sag11}
E. Saglamyurek, N. Sinclair, J. Jin, J. A. Slater, D. Oblak,  F. Bussi\`eres, M. George, R. Ricken, W. Sohler and W. Tittel,
\newblock Nature \textbf{469}, 512 (2011). 

\bibitem{clausen11}
C. Clausen, I. Usmani, F. Bussi\`eres, N. Sangouard, M. Afzelius, H. de Riedmatten, N. Gisin,
\newblock Nature \textbf{469}, 508 (2011).

\bibitem{Damon10}
V. Damon, M. Bonarota, A. Louchet-Chauvet, T. Chaneli\`ere, J.-L. Le Gou\"et,
\newblock New Journal of Physics, \textbf{13}, 093031 (2011).


\bibitem{GuillotNoel2006}
O. Guillot-No\"{e}l, Ph. Goldner, Y. Le Du, E. Baldit, P.Monnier, and K. Bencheikh,
\newblock Phys. Rev. B \textbf{74}, 214409 (2006).

\bibitem{Diniz2011}
I. Diniz, S. Portolan, R. Ferreira, J.M. Gerard, P. Bertet, and A. Auffeves,
\newblock Phys. Rev. A \textbf{84}, 063810 (2011).


\bibitem{Sun08}
Y. Sun, T. B\"ottger, C. W. Thiel, R. L. Cone,
\newblock Phys. Rev. B \textbf{77}, 085124 (2008).

\bibitem{bottger02}
T. B\"ottger,
\newblock PhD thesis, Montana State University, Bozeman, Montana, USA (2002).

\bibitem{schuster10}
D. I. Schuster, A.P. Sears, E. Ginossar, L. DiCarlo, L. Frunzio, J.J.L. Morton, H. Wu, G.A.D. Briggs, B.B. Buckley, D.D. Awschalom, R.J. Schoelkopf, 
\newblock Phys. Rev. Lett., \textbf{105} 140501 (2011).


\bibitem{hyperfine}
The hyperfine site peaks are detuned by: $\Delta_{\text{hyper}\pm{}}=\omega_r -g_l\mu_b(B\pm{A_0M_I})/\hbar $ where A$_0$ is the hyperfine coupling constant and M$_I$ is the nuclear spin projection.





\end{thebibliography}

\end{document}